# Thermal Properties of Deng-Fan-Eckart Potential model using Poisson Summation Approach

C.O.Edet[1*], U.S.Okorie[1&2], G.Osobonye[3], A.N.Ikot[1], G.J.Rampho[4] and R.Sever[5].

[1]Theoretical Physics Group, Department of Physics, University of Port Harcourt, Port Harcourt-Nigeria.

[2]Department of Physics, Akwa Ibom State University, Ikot Akpaden, Uyo.-Nigeria

[3]Department of Physics, Federal College of Education, Omuku, Rivers State- Nigeria

[4]Department of Physics, University of South Africa.

[5]Department of Physics, Middle East Technical University, 06800, Ankara, Turkey.

## ABSTRACT

The Deng-Fan-Eckart (DFE) potential is as good as the Morse potential in studying atomic interaction in diatomic molecules. By using the improved Pekeris-type approximation, to deal with the centrifugal term, we obtain the bound-state solutions of the radial Schrödinger equation with this adopted molecular model via the Factorization Method. With the energy equation obtained, the thermodynamic properties of some selected diatomic molecules ($H_2$, $CO$, $ScN$ and $ScF$) were obtained using Poisson summation method.. The unnormalized wave function is also derived. The energy spectrum for a set of diatomic molecules for different values of the vibrational $n$ and rotational $\ell$ are obtained. To show the accuracy of our results, we discuss some special cases by adjusting some potential parameters and also compute the numerical eigenvalue of the Deng-Fan potential for comparison sake. However, it was found out that our results agree excellently with the results obtained via other methods.

*corresponding author email:collinsokonedet@gmail.com

# 1. INTRODUCTION

An adequate understanding of molecular behavior of two interacting atoms as a function of their relative positions is very pertinent in different fields of Physics and Chemistry. Recently, there has been a great interest in obtaining molecular potential energy functions governing the interaction of two atom (diatomic molecules) [1-5].

In the available literature, numerous efforts have been made by researchers to construct a quasi-precise analytical molecular potential energy function [6-10]. Quite recently, such parameters like the dissociation energy and equilibrium bond length diatomic molecules as explicit parameters has been harnessed by some researchers[11] to construct improved and modified versions for some empirical potential energy functions with more applications, this functions includes; the well-known Rosen-Morse[12], Tietz[13], Frost-Musulin[14], Manning-Rosen[15], Poschl-Teller[16] and Deng-Fan[17] potentials. These authors found that the well-known Tietz potential function is conventionally defined in terms of five parameters but it actually has only four independent parameters. Following the same path, Fu et al. [6] demonstrated that the five-parameter exponential-type potential is identical to the Tietz potential for diatomic molecules. It was observed that the improved five-parameter exponential-type potential can well model the inter-nuclear interaction potential energy curve for the ground electronic state of the carbon monoxide molecule by the utilization of the experimental values of three molecular constants

These modified and or improved molecular potentials have received remarkable attentions. This is due to their various applications in many fields of Physics and Chemistry. Diatomic molecular potentials has been applied to simulate molecular potential energy curves, predict thermo-chemical properties of diatomic molecules[18], calculations of molecular vibrational partition function [19-21] and prediction of enthalpy and entropy of gaseous dimers[22-26].

Berkdemir et al. [27] proposed an improved version of the Kratzer potential[28], Hamzavi et al.[30] proposed an improved expression for the Deng-Fan oscillator potential[17], it was found out that the shifted Deng-Fan was more Morse-like when compared on a plot than the original Deng-Fan oscillator(see. Fig.1 of Ref[30]).The same route was taken by Falaye et al [32] to propose the shifted Tietz–Wei (STW) oscillator. They shifted the well-known Tietz-Wei potential [33-36] by $D_e$. In their presentation, it was noted that STW is as good as the traditional Morse potential [3, 37] in simulating the atomic interaction in diatomic molecules.

Several methods have been adopted to study these potentials in the relativistic and non-relativistic regime, these includes the Nikiforov Uvarov method[38,39], Factorization Method[40], Modified Factorization Method[20, 21],Asymptotic Iteration Method(AIM)[41, 42] and Supersymmetric Quantum Mechanics(SUSYQM)[43, 44], Exact Quantization Rule[45-49], Proper quantization Rule[50-54] and Formula Method[55] amongst others.

These great attempts were made in a quest to find the most suitable molecular potential for the description of diatomic molecules. Motivated by works in Refs. [27, 30 and 32], we adopt the Deng-Fan-Eckart Potential(DFEP) proposed by Ikot et al. [59] to study interatomic interactions of diatomic molecules .This potential is given by[56]

$$V(r) = D_e \left(1 - \frac{b}{(e^{\alpha r} - 1)}\right)^2 - \frac{V_1 e^{-\alpha r}}{(1 - e^{-\alpha r})} + \frac{V_2 e^{-\alpha r}}{(1 - e^{-\alpha r})^2} \tag{1}$$

$$b = e^{\alpha r_e} - 1$$

where $r_e$ is the molecular bond length, $D_e$ is the dissociation energy, $r$ is the inter-nuclear distance, $\alpha$ the range of the potential well, $V_1$ and $V_2$ are the potential strengths. This potential is just a superposition of the Deng-Fan and Eckart potentials. The shape of this potential is shown in Fig. 1.

Ikot et al [56] solved the Dirac equation under spin and pseudo-spin symmetries Deng–Fan-Eckart potential with Coulomb-like and Yukawa-like tensor interaction terms. The energy equation was obtained by using the Nikiforov–Uvarov method. This study is an extension of the work of ref. [56].

The thermal properties for the Deng–Fan-Eckart potential are also studied in this work. These include vibrational partition function, vibrational mean energy, vibrational mean free energy, vibrational entropy, etc. Prior to now, different authors have investigated the thermodynamic properties for some quantum mechanical systems [19-22].

The objective of this study is in three fold; firstly, we solve the Schrodinger with the DFE potential using the Factorization method, we apply the energy equation obtained to study atomic behavior in some selected diatomic molecules and also study the thermodynamic properties of this potential for some selected molecules.

The scheme of our research article is as follows. In the next section, we solve the radial Schrödinger equation with the Deng-Fan-Eckart Potential using the Factorization method [40] and in section 3, we obtain the thermodynamic properties for the diatomic molecules which will be calculated using the expression for the partition function. In section 4, we obtain the rotational-vibrational energy spectrum for some diatomic molecules with numerical results and discussion. In section 5, we present special case of the potential under consideration. Finally, in section 6 we give concluding remark.

## 2. Energy levels and Wavefunction

The radial part of the Schrödinger equation is given by[57];

$$\frac{d^2 \psi(r)}{dr^2} + \frac{2\mu}{\hbar^2}\left[E_{n\ell} - V(r) - \frac{\hbar^2 \ell(\ell+1)}{2\mu r^2}\right]\psi(r) = 0 \tag{2}$$

Considering Deng-Fan-Eckart potential (Eq.(1)), we obtain the radial Schrödinger equation, Eq.(2) is rewritten as follows:

$$\frac{d^2\psi(r)}{dr^2}+\frac{2\mu}{\hbar^2}\left[\begin{array}{c}E_{n\ell}-\left(D_e+\frac{D_e b^2 e^{-2\alpha r}}{(1-e^{-\alpha r})^2}-\frac{2D_e b e^{-\alpha r}}{(1-e^{-\alpha r})}-\frac{V_1 e^{-\alpha r}}{(1-e^{-\alpha r})}+\frac{V_2 e^{-\alpha r}}{(1-e^{-\alpha r})^2}\right)\\ -\frac{\hbar^2 \ell(\ell+1)}{2\mu r^2}\end{array}\right]\psi(r)=0 \quad (3)$$

This equation cannot be solved analytically for $\ell \neq 0$ due to the centrifugal term. Therefore, we must use an approximation to the centrifugal term. We use the following approximation [30]

$$\frac{1}{r^2} \approx \alpha^2 \left[d_0 + \frac{e^{\alpha r}}{(1-e^{\alpha r})^2}\right] \quad (4)$$

It has been shown extensively in literature that this approximation scheme approximates the centrifugal barrier better than the well-known Greene and Aldrich[58] approximation scheme[31]. In the limiting case when $\alpha r \to 1$, the value of the dimensionless constant $d_0 = \frac{1}{12}$ and the screening parameter $\alpha$ approaches zero, Eq.(4) reduces to $\frac{1}{r^2}$.

Inserting Eqs. (4) into Eq. (3) where we have the $\frac{1}{r^2}$ term, we have;

$$\frac{d^2\psi(r)}{dr^2}+\frac{2\mu}{\hbar^2}\left[\begin{array}{c}E_{n\ell}-\left(D_e+\frac{D_e b^2 e^{-2\alpha r}}{(1-e^{-\alpha r})^2}-\frac{2D_e b e^{-\alpha r}}{(1-e^{-\alpha r})}-\frac{V_1 e^{-\alpha r}}{(1-e^{-\alpha r})}+\frac{V_2 e^{-\alpha r}}{(1-e^{-\alpha r})^2}\right)-\\ \frac{\hbar^2 \ell(\ell+1)\alpha^2}{2\mu}\left(d_0+\frac{e^{\alpha r}}{(1-e^{\alpha r})^2}\right)\end{array}\right]\psi(r)=0 \quad (5)$$

For Mathematical simplicity, let's introduce the following dimensionless notations;

$$-\varepsilon^2 = \frac{2\mu(E_{n\ell}-D_e)}{\hbar^2 \alpha^2}, \eta_1 = \frac{2\mu D_e b^2}{\hbar^2 \alpha^2},$$

$$\eta_2 = \frac{4\mu D_e b}{\hbar^2 \alpha^2}, \beta_1 = \frac{2\mu V_1}{\hbar^2 \alpha^2}, \beta_2 = \frac{2\mu V_2}{\hbar^2 \alpha^2}$$

$$\gamma = \ell(\ell+1). \quad (6)$$

Hence, Eq.(5) can be rewritten in a less ambiguous form as;

$$\frac{d^2\psi(r)}{dr^2}+\left[\begin{array}{c}-\varepsilon^2\left(1-e^{-\alpha r}\right)^2-\eta_1 e^{-2\alpha r}+\eta_2 e^{-\alpha r}\left(1-e^{-\alpha r}\right)+\beta_1 e^{-\alpha r}\left(1-e^{-\alpha r}\right)\\ -\beta_2 e^{-\alpha r}-\gamma e^{-\alpha r}-\gamma d_0 \left(1-e^{-\alpha r}\right)^2\end{array}\right]\psi(r)=0 \quad (7)$$

Using the coordinate transformation $s = e^{-\alpha r}$, Eq.(7) translates to;

$$\frac{d^2\psi(s)}{ds^2}+\frac{1}{s}\frac{d\psi(s)}{ds}+\frac{1}{s^2(1-s)^2}\left[\begin{array}{c}-(\varepsilon^2+\eta_1+\eta_2+\beta_1+\gamma d_0)s^2\\ +(2\varepsilon^2+\eta_2+\beta_1-\beta_2-\gamma+2\gamma d_0)s-(\varepsilon^2+\gamma d_0)\end{array}\right]\psi(s)=0 \quad (8)$$

If we consider the boundary conditions

$$s \Rightarrow \begin{cases} 0, & r \to \infty, \\ 1, & r \to 0, \end{cases} \qquad (9)$$

with $\psi(s) \to 0$, we take the following radial wave functions of the form

$$\psi(s) = s^\lambda (1-s)^\omega f(s) \qquad (10)$$

where

$$\lambda = \sqrt{\varepsilon^2 + \gamma d_0} \qquad (11)$$

$$\omega = \frac{1}{2} + \sqrt{\frac{1}{4} + \eta_1 + \beta_2 + \gamma} \qquad (12)$$

On substitution of Eq. (10) into Eq. (8) leads to the following hypergeometric equation:

$$s(1-s)\frac{d^2 f(s)}{ds^2} + \left[(2\lambda+1) - (2\lambda+2\omega+1)s\right]\frac{df(s)}{ds} - \left[\frac{(\lambda+\omega)^2}{-(\varepsilon^2 + \eta_1 + \eta_2 + \beta_1 + \gamma d_0)}\right]f(s) = 0 \qquad (13)$$

whose solutions are nothing but the hypergeometric functions

$$f(s) = {}_2F_1(a,b;c;s) \qquad (14)$$

where

$$a = (\lambda+\omega) - \sqrt{\varepsilon^2 + \eta_1 + \eta_2 + \beta_1 + \gamma d_0}$$
$$b = (\lambda+\omega) + \sqrt{\varepsilon^2 + \eta_1 + \eta_2 + \beta_1 + \gamma d_0} \qquad (15)$$
$$c = 2\lambda + 1$$

By considering the finiteness of the solutions, the quantum condition is given by

$$(\lambda+\omega) - \sqrt{\varepsilon^2 + \eta_1 + \eta_2 + \beta_1 + \gamma d_0} = -n, \qquad n = 0,1,2... \qquad (16)$$

from which we obtain

$$\varepsilon^2 = -\gamma d_0 + \frac{1}{4}\left(\frac{\left(n + \frac{1}{2} + \sqrt{\frac{1}{4} + \eta_1 + \beta_2 + \gamma}\right)^2 - \eta_1 - \eta_2 - \beta_1}{\left(n + \frac{1}{2} + \sqrt{\frac{1}{4} + \eta_1 + \beta_2 + \gamma}\right)}\right)^2 \qquad (17)$$

Thus, if one substitutes the value of the dimensionless parameters in Eq.(6) into Eq.(17), we obtain the solutions as follows:

$$E_{n\ell} = D_e + \frac{\hbar^2 \alpha^2 \ell(\ell+1)d_0}{2\mu} - \frac{\hbar^2 \alpha^2}{2\mu}\left(\frac{(n+\nu)^2 - \frac{2\mu D_e b^2}{\hbar^2 \alpha^2} - \frac{4\mu D_e b}{\hbar^2 \alpha^2} - \frac{2\mu V_1}{\hbar^2 \alpha^2}}{2(n+\nu)}\right)^2 \qquad (18)$$

Where,

$$\nu = \frac{1}{2} + \frac{1}{2}\sqrt{(2\ell+1)^2 + \frac{8\mu D_e b^2}{\hbar^2 \alpha^2} + \frac{8\mu V_2}{\hbar^2 \alpha^2}} \qquad (19)$$

The corresponding unnormalized wave function is obtain as

$$\psi(s) = N_{n\ell} s^{\sqrt{\varepsilon^2 + \gamma d_0}} (1-s)^{\frac{1}{2}+\sqrt{\frac{1}{4}+\eta_1+\beta_2+\gamma}} {}_2F_1\left(-n, n+2(\lambda+\omega); 2\lambda+1; s\right) \qquad (20)$$

## 3. THERMAL PROPERTIES OF DFEP

The vibrational partition function can be calculated with the aid of direct summation over all possible vibrational energy levels at a given temperature $T$ to be [19]

$$Z(\beta) = \sum_{n=0}^{n_{max}} e^{-\beta E_{nl}}, \quad \beta = \frac{1}{k_B T} \qquad (21)$$

Here, $k_B$ is the Boltzmann constant and $E_{nl}$ is the rotational-Vibrational energy of the nth bound state.

We can rewrite eq. (18) to be of the form

$$E_{nl} = P_1 - \frac{\hbar^2 \alpha^2}{2\mu} \left[ \frac{(n+v)}{2} + \frac{P_2}{2(n+v)} \right]^2 \qquad (22)$$

where

$$P_1 = D_e + \frac{\hbar^2 \alpha^2 l(l+1) d_0}{2\mu}; \quad P_2 = -\left( \frac{2\mu D_e b^2}{\hbar^2 \alpha^2} + \frac{4\mu D_e b}{\hbar^2 \alpha^2} + \frac{2\mu V_1}{\hbar^2 \alpha^2} \right) \qquad (23)$$

We substitute eq. (22) into eq. (21) to have

$$Z(\beta) = \sum_{n=0}^{n_{max}} e^{-\beta \left[ P_1 - \frac{\hbar^2 \alpha^2}{2\mu} \left( \frac{(n+v)}{2} + \frac{P_2}{2(n+v)} \right)^2 \right]} \qquad (24)$$

where

$$n_{max} = \sqrt{P_2} - v \qquad (25)$$

We adopt the Poisson summation formula given as [59]

$$\sum_{n=0}^{n_{max}} f(x) = \frac{1}{2}\left[ f(0) - f(n_{max}+1) \right] + \sum_{m=-\infty}^{\infty} \int_0^{n_{max}+1} f(y) e^{-i 2\pi m y} dy \qquad (26)$$

But, for the lower order approximation, the Poisson summation formula reduces to [59]

$$\sum_{n=0}^{n_{max}} f(x) = \frac{1}{2}\left[ f(0) - f(n_{max}+1) \right] + \int_0^{n_{max}+1} f(y) dy \qquad (27)$$

By the application of eq. (27) for the partition function of eq. (24), we obtain

$$Z(\beta) = \frac{1}{2}\left[ e^{-\beta(P_1 - q_1 q_2^2)} - e^{-\beta(P_1 - q_1 q_3^2)} + \int_0^{n_{max}} \left( e^{-L_1 \beta - \frac{L_2 \beta}{\rho^2} - L_3 \beta \rho^2} \right) d\rho \right] \quad (28)$$

where

$$q_1 = \frac{\hbar^2 \alpha^2}{2\mu}; \quad q_2 = \left( \frac{P_2}{2 n_{max}} + \frac{n_{max}}{2} \right); \quad q_3 = \left( \frac{P_2}{2(n_{max} + 1 + v)} + \frac{(n_{max} + 1 + v)}{2} \right);$$

$$\rho = y + v; \quad L_1 = \frac{\hbar^2 \alpha^2 P_2^2}{4\mu} - P_1; \quad L_2 = \frac{\hbar^2 \alpha^2 P_2^2}{8\mu}; \quad L_3 = \frac{\hbar^2 \alpha^2}{4\mu} \quad (29)$$

We therefore use the Maple software to evaluate the integral in eq. (36), thus obtaining the Poisson summation rotational-vibrational partition function for the diatomic molecules with DFEP models as

$$Z(\beta) = \frac{1}{2}\left[ e^{-\beta(P_1 - q_1 q_2^2)} - e^{-\beta(P_1 - q_1 q_3^2)} + e^{-L_3 \beta \rho^2 - L_1 \beta} \sqrt{L_2 \beta} \left( \frac{2 n_{max} e^{-\frac{L_2 \beta}{n_{max}^2}}}{\sqrt{L_2 \beta}} + \frac{2\sqrt{L_2 \beta}\sqrt{\pi}\, erf\left(\frac{\sqrt{L_2 \beta}}{n_{max}}\right)}{\sqrt{L_2 \beta}} - 2\sqrt{\pi} \right) \right] \quad (30)$$

The imaginary error function can be defined as [21]

$$erfi(z) = ierf(z) = \frac{2}{\sqrt{\pi}} \int_0^z e^{u^2} du \quad (31)$$

Thermodynamic functions such as Helmholtz free energy, $F(\beta)$, entropy, $S(\beta)$, internal energy, $U(\beta)$, and specific heat, $C_v(\beta)$, functions can be obtained from the partition function (30) as follows [20];

$$F(\beta) = -\frac{1}{\beta} \ln Z(\beta) \quad (32)$$

$$S(\beta) = -k_\beta \frac{\partial F(\beta)}{\partial \beta} \quad (33)$$

$$U(\beta) = -\frac{\partial (\ln Z(\beta))}{\partial \beta} \quad (34)$$

$$C_v = k_\beta \frac{\partial U(\beta)}{\partial \beta} \quad (35)$$

## 4. Applications

To show the accuracy of energy equation obtained in this work, we calculate the energy levels using Eq.(18) for different quantum numbers $n$ and $\ell$ with for four diatomic molecules ($H_2$, $CO$, $ScN$ and $ScF$). In our numerical computations, we have used spectroscopic parameters shown in Table 1 and the following conversions: $1\,amu = 931.494028\,MeV/c^2$, $1cm^{-1} = 1.239841875 \times 10^{-4}\,eV$ and $\hbar c = 1973.29\,eV\,\text{Å}$.

The numerical values of these energies for different vibrational and rotational quantum numbers are presented in Table 4. To further validate our results, we have also computed the energy eigenvalues of the Deng-Fan potential using the reduced energy equation given in Eq. (22) as a special case. Our results shown in Tables 2 and 3 are in good agreement with the results given in Ref. [31].

Figs. 1 shows a comparison between DFE diatomic molecular potential, and the Morse potential using the parameters set for $H_2$ diatomic molecule. As it can be seen from this plot, the Deng-Fan-Eckart and the Morse potentials are very close to each other for large values of $r$ in the regions $r \approx r_e$ and $r > r_e$, but they are very different at $r = 0$. This implies that the Deng-Fan-Eckart potential is as good as the Morse potential in simulating the molecular interaction for diatomic molecules. The shape of this potential is shown in Fig. 1 for different molecules. Fig. 3 shows the energy eigenvalues variation with dissociation energy for various vibrational quantum states. It can be easily observed from Fig. (3) that the dissociation energy increases directly as the energy increases. Fig. 4 shows the energy eigenvalues variation with equilibrium bond length for various vibrational quantum numbers. In Fig. 4, the minimum value of the energy is observed when the value of $r_e$ is in the region of $0.1 - 0.2\,\text{Å}$. Beyond this region, the energy begins to decrease abruptly. More so, an asymptotic behavior is observed in the curve representing the 2p state. Fig. 5 shows the energy eigenvalues variation with screening parameter for various vibrational quantum numbers. From the plot, it can be seen that there is a uniform convergence of the energy curves at the point where $\alpha \approx 0.1$ (i.e. in the low screening region) and $E_{n\ell} \approx 50\,eV$. Beyond this region, the curves uniformly diverge and there is a sharp decrease in energy $\alpha$ increases except in the 2p state, where the energy was almost constant. Figure 6 shows Energy eigenvalues variation with parameter $V_1$ for various vibrational quantum numbers. All curves representing the energy states show a uniform decrease in energy as $V_1$ increases. In Fig. 7 shows the energy eigenvalues variation with parameter $V_2$ for various vibrational quantum numbers and the energy increases monotonically with $V_2$.

Figure 8 shows the energy eigenvalues variation with dissociation energy for various diatomic molecules. It is seen that the dissociation energy increases directly as the energy increases for all diatomic molecules except for $ScN$ molecule which shows an almost-linear trend. Figure 9 shows the variation of the energy of the system with the equilibrium bond length for the four diatomic molecules understudy. It's seen that in the low region of $r_e$, the energy drops to almost zero. Beyond $r_e \approx 0$, we notice a sharp rise and this continues in a quasi-linear manner. Figure 10 shows the energy variation with the screening parameter $\alpha$. In the low screening region ($\alpha \approx 0-1$), we observe a sharp rise in the energy, beyond this point, there is a sharp decrease in the energy eigenvalue and this continues as the screening parameter increase. For $ScF$ molecule curve, we observe a quasi-asymptotic behavior. Figures 11 and 12 shows the variation of the energy of the system with parameters $V_1$ and $V_2$ for different molecules studied. It is seen that a linear relationship exist between the energy and the parameters. Figure 13 explicitly shows the energy eigenvalues variation with the particle mass $\mu$ for various diatomic molecules. It can be seen that in the region $\mu \approx 0-0.1\, a.m.u$, the energy eigenvalue drops sporadically. This continues in a linear trend. The energy is only high in the region where the mass is very low but decreases as the particle's mass increases. The energy is very similar for $0.2 < \mu < 1.0$ .

In Fig. 14 we show the vibrational partition function variation with temperature for various diatomic molecules. It can be seen that the partition function is almost invariant with increasing Temperature for the $H_2$ molecule. For the other three molecules, we observed that in the region $\beta \approx 0-1\times 10^{-12}$, the partition function is very low, beyond this point, it rises abruptly. Fig. 15 shows the Vibrational free energy variation with temperature for various diatomic molecules. It is seen that the vibrational free energy increases monotonically with increasing Temperature. Fig. 16 and 17 shows the Vibrational mean energy and entropy variation with temperature for various diatomic molecules. Again, it can be seen that the Vibrational mean energy and entropy is almost invariant with increasing Temperature for the $H_2$ molecule. For other three molecules considered, the Vibrational mean energy and entropy decreases as the Temperature increases. Fig. 18 shows the vibrational specific heat capacity variation with temperature for various diatomic molecules. In like manner, the vibrational specific heat capacity is invariant with increasing Temperature for the $H_2$ molecule. For other three molecules, we observe that the vibrational specific heat capacity increases sporadically as the temperature increases.

## 5. Special cases

In this section, we make some adjustments of constants in Eq. (1) and Eq. (18) to have the following cases:

**Deng-Fan Potential**

If we set $V_1$ and $V_2$ equal to zero, the DFE potential model (Eq. 1) reduces to the Deng-Fan potential

$$V(r) = D_e\left(1 - \frac{b}{(e^{\alpha r} - 1)}\right)^2 \qquad (36)$$

$$b = e^{\alpha r_e} - 1$$

In like manner, Eq.(18) reduces to the energy equation for the Deng-Fan potential

$$E_{n\ell} = D_e + \frac{\hbar^2 \alpha^2 \ell(\ell+1)d_0}{2\mu} - \frac{\hbar^2 \alpha^2}{2\mu}\left(\frac{\left(n + \frac{1}{2} + \frac{1}{2}\sqrt{(2\ell+1)^2 + \frac{8\mu D_e b^2}{\hbar^2 \alpha^2}}\right)^2 - \frac{2\mu D_e b^2}{\hbar^2 \alpha^2} - \frac{4\mu D_e b}{\hbar^2 \alpha^2}}{\left(2n + 1 + \sqrt{(2\ell+1)^2 + \frac{8\mu D_e b^2}{\hbar^2 \alpha^2}}\right)}\right)^2 \qquad (37)$$

Eq.(37) is in agreement with Eq. (29) of ref.[31] who solved the Schrodinger equation with this potential using the NU method with the same approximation. If $d_0 = 0$, the energy reduces to the result obtained in Eq.(20) of ref.[60].

**Eckart Potential**

If we set $D_e$ equal to zero, the DFE potential model (Eq. 1) reduces to the Deng-Fan potential

$$V(r) = -\frac{V_1 e^{-\alpha r}}{(1-e^{-\alpha r})} + \frac{V_2 e^{-\alpha r}}{(1-e^{-\alpha r})^2} \tag{38}$$

$$E_{n\ell} = \frac{\hbar^2 \alpha^2 \ell(\ell+1)d_0}{2\mu} - \frac{\hbar^2 \alpha^2}{2\mu} \left( \frac{\left(n+\frac{1}{2}+\frac{1}{2}\sqrt{(2\ell+1)^2 + \frac{8\mu V_2}{\hbar^2 \alpha^2}}\right)^2 - \frac{2\mu V_1}{\hbar^2 \alpha^2}}{2n+1+\sqrt{(2\ell+1)^2 + \frac{8\mu V_2}{\hbar^2 \alpha^2}}} \right)^2 \tag{39}$$

If $d_0 = 0$, Eq.(39) reduces to the energy equation obtained in Eq.(31) of ref.[61].

**Hulthen Potential**

If we set $D_e$ and $V_2$ equal to zero, the DFE potential model (Eq. 1) reduces to the Deng-Fan potential

$$V(r) = -\frac{V_1 e^{-\alpha r}}{(1-e^{-\alpha r})} \tag{40}$$

$$E_{n\ell} = \frac{\hbar^2 \alpha^2 \ell(\ell+1)d_0}{2\mu} - \frac{\hbar^2 \alpha^2}{2\mu} \left( \frac{(n+\ell+1)^2 - \frac{2\mu V_1}{\hbar^2 \alpha^2}}{2(n+\ell+1)} \right)^2 \tag{41}$$

Eq.(41) is identical with the energy eigenvalues formula given in Eq. (36) of ref.[62],Eq. (32) of ref.[63], Eq. (24) of ref.[64] and Eq. (28) of [65].

### 6. Conclusion

In this article, we have solved the Schrodinger equation using factorization method and suitable approximation to overcome the centrifugal barrier. We have expressed the solutions by the generalized hypergeometric functions $_2F_1(a,b;c;s)$. We have also presented the ro-vibrational energy spectra with the Deng-Fan-Eckart potential model for some diatomic molecules. Computation of energies have been done numerically and the results discussed extensively using plots. In detail, we evaluated the vibrational partition functions $Z(\beta)$ which we used to study the thermodynamics properties of vibrational mean energy $U(\beta)$, vibrational entropy $S(\beta)$, vibrational mean free energy $F(\beta)$ and vibrational specific heat capacity $C(\beta)$ for some selected diatomic molecules. We discussed some special cases by adjusting the potential parameters and compute the numerical energy spectra for the Deng-Fan potential. It was found that our results agree with the existing literature. More so, we also found out that the Deng-Fan-Eckart potential can effectively describe the vibrational energy levels of diatomic molecules.

Recently, there has been investigations by some researchers reporting some interesting description of internal vibrations of diatomic molecules. Worthy to point out are the works that have successfully predicted the thermodynamic properties for some diatomic gases and gaseous dimmers, including CO, $N_2$, $Cl_2$ and gaseous sodium dimer and lithium dimer [68-74]. Finally, this study has many applications in different areas of physics and chemistry such as atomic physics, molecular physics and chemistry amongst others

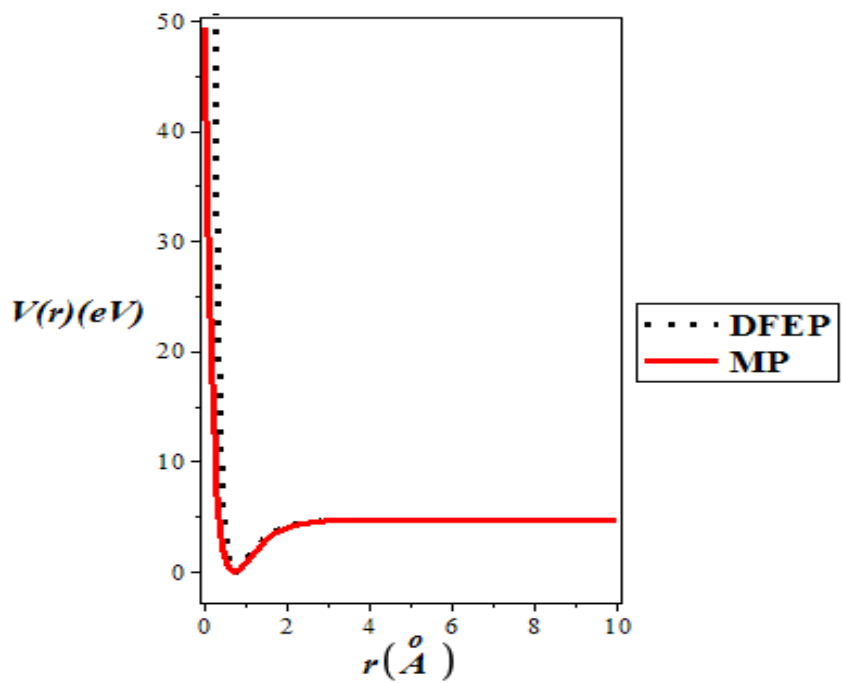

Figure 1: Shape of the Deng-Fan-Eckart potential model and Morse oscillator potentials for $H_2$ diatomic molecule.

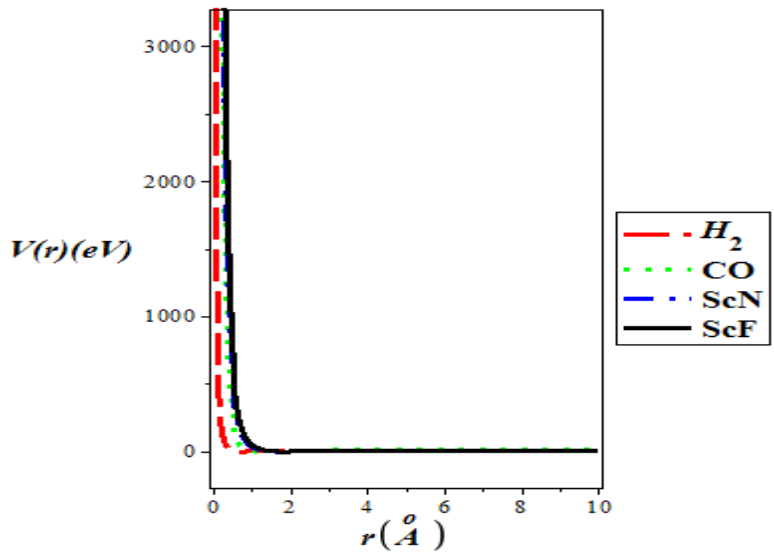

Figure 2: Shape of the Deng-Fan-Eckart diatomic molecular potential for different diatomic molecules.

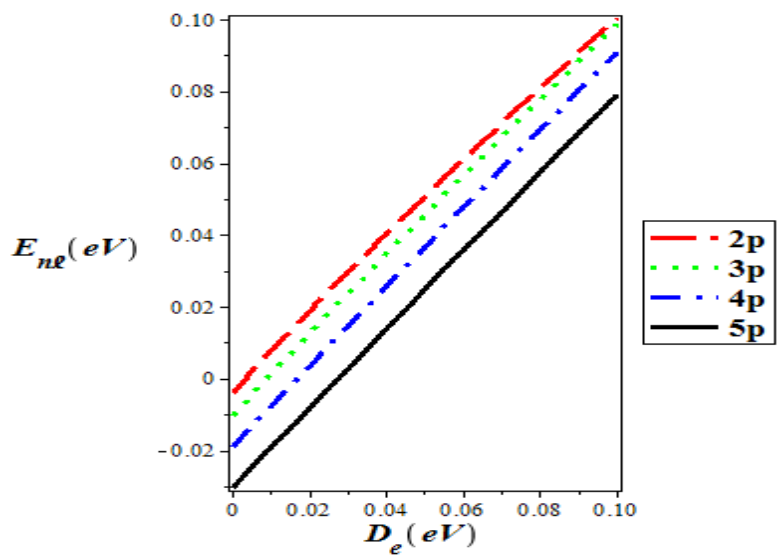

Figure 3: energy eigenvalues variation with dissociation energy for various quantum state

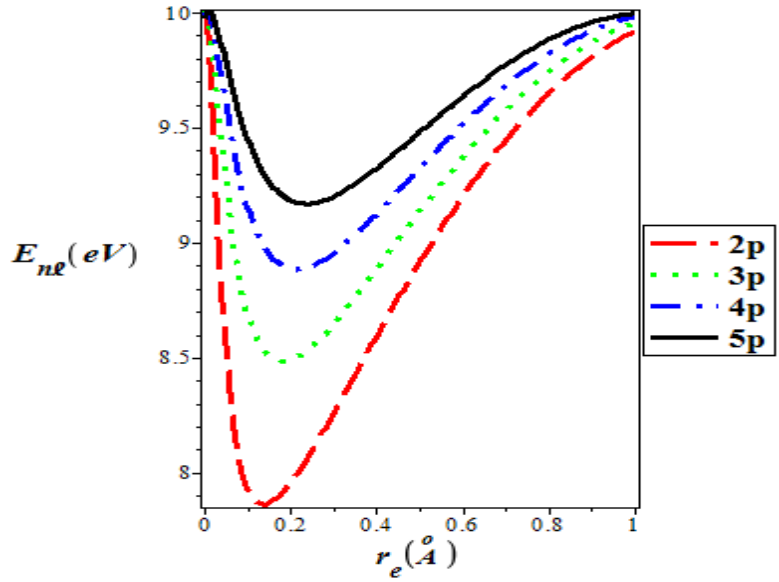

Figure 4: energy eigenvalues variation with equilibrium bond length for various quantum states

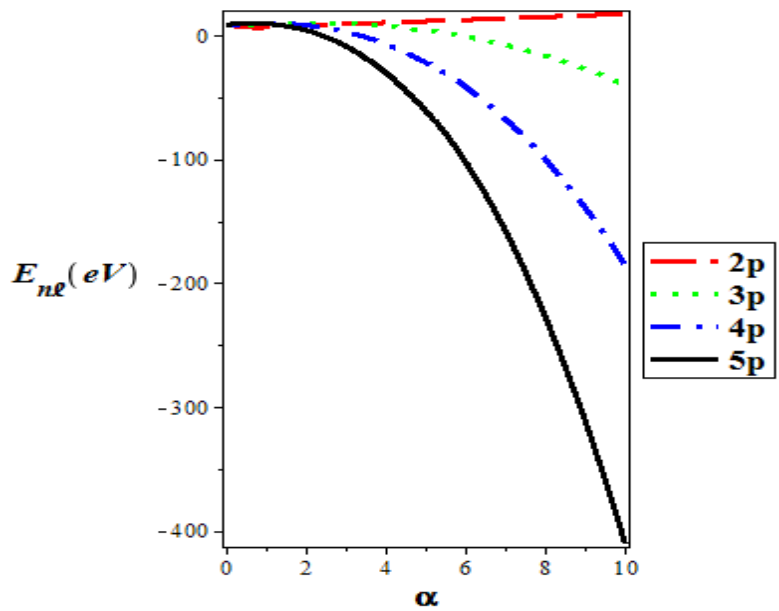

Figure 5: energy eigenvalues variation with screening parameter for various quantum states

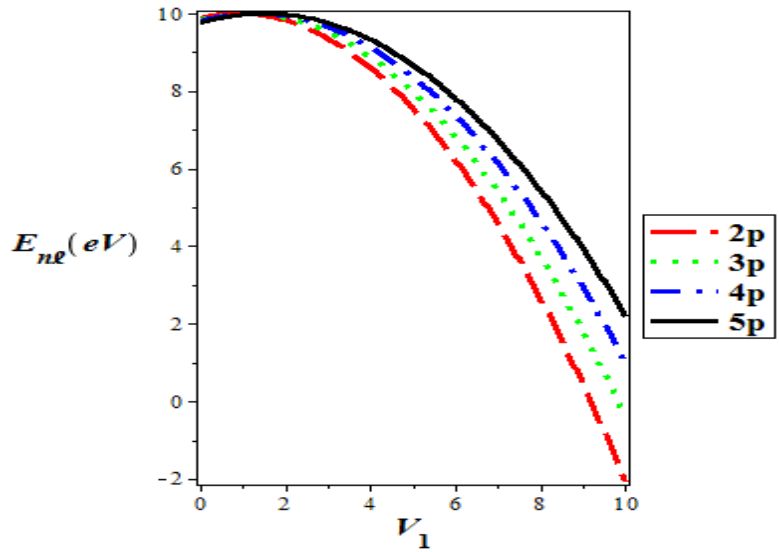

Figure 6: energy eigenvalues variation with parameter $V_1$ for various quantum states

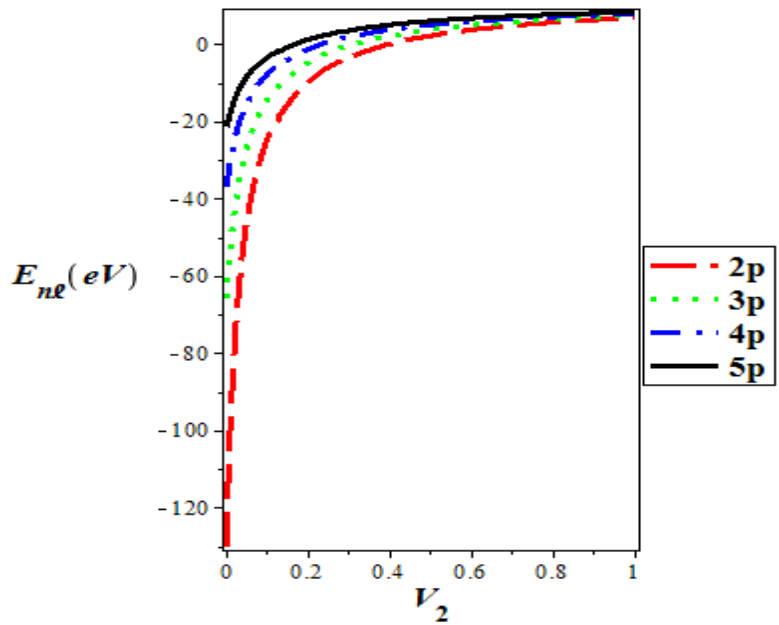

Figure 7: energy eigenvalues variation with parameter $V_2$ for various quantum states

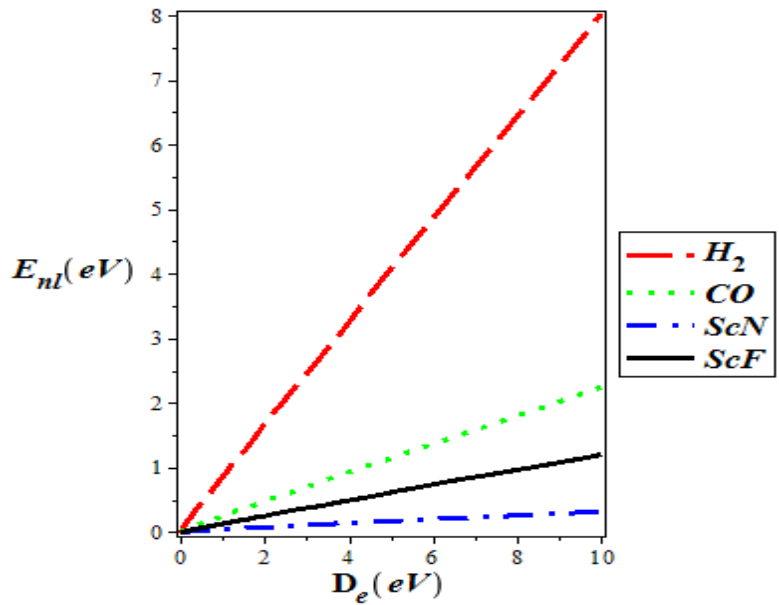

Figure 8: energy eigenvalues variation with dissociation energy for various diatomic molecules

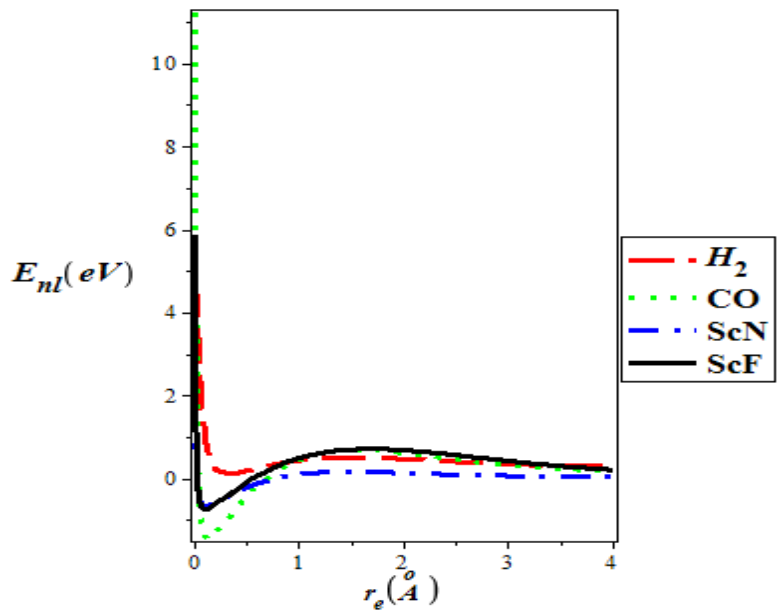

Figure 9: energy eigenvalues variation with equilibrium bond length for various diatomic molecules

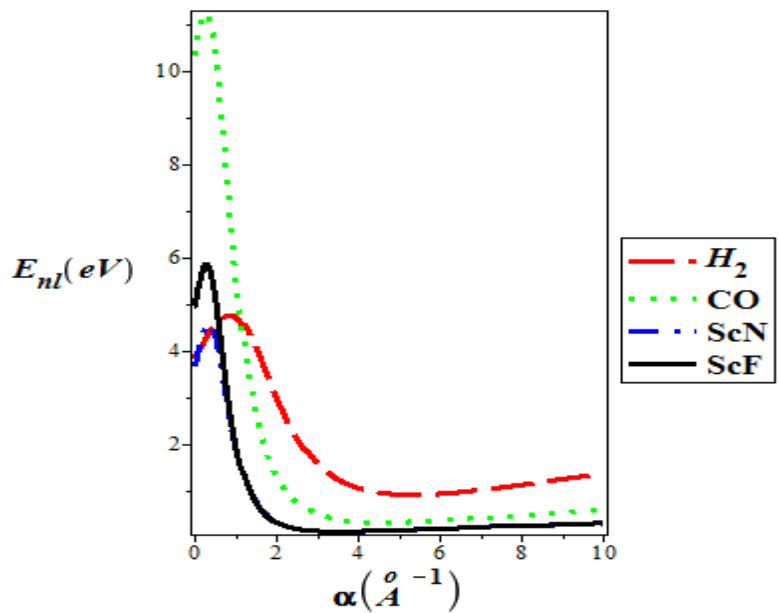

Figure 10: energy eigenvalues variation with screening parameter for various diatomic molecules

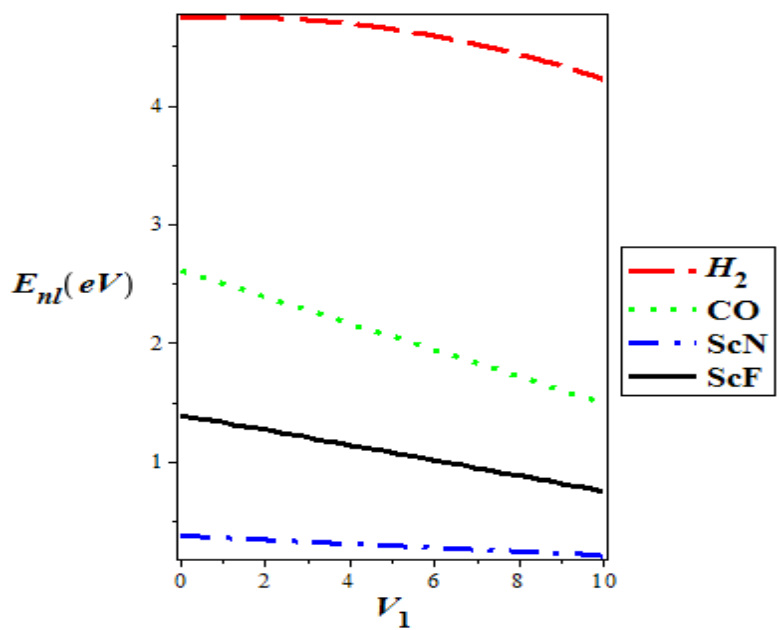

Figure 11: energy eigenvalues variation with the parameter $V_1$ for various diatomic molecules

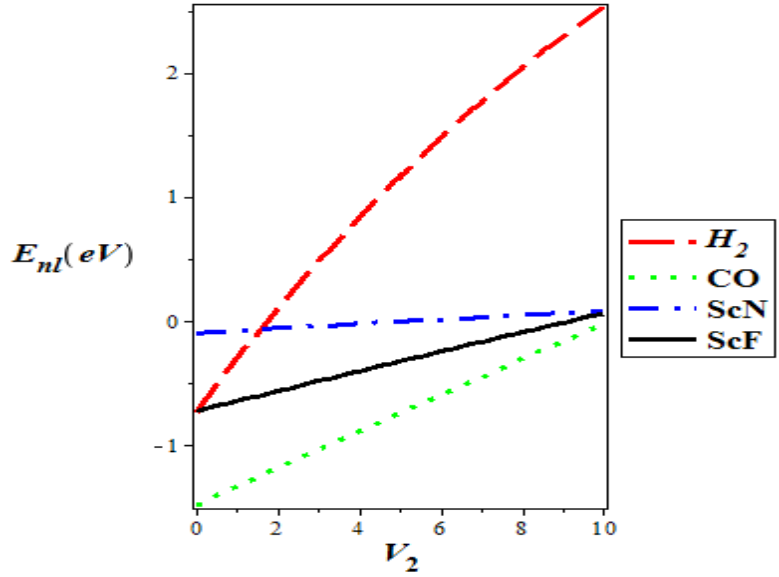

Figure 12: energy eigenvalues variation with the parameter $V_2$ for various diatomic molecules

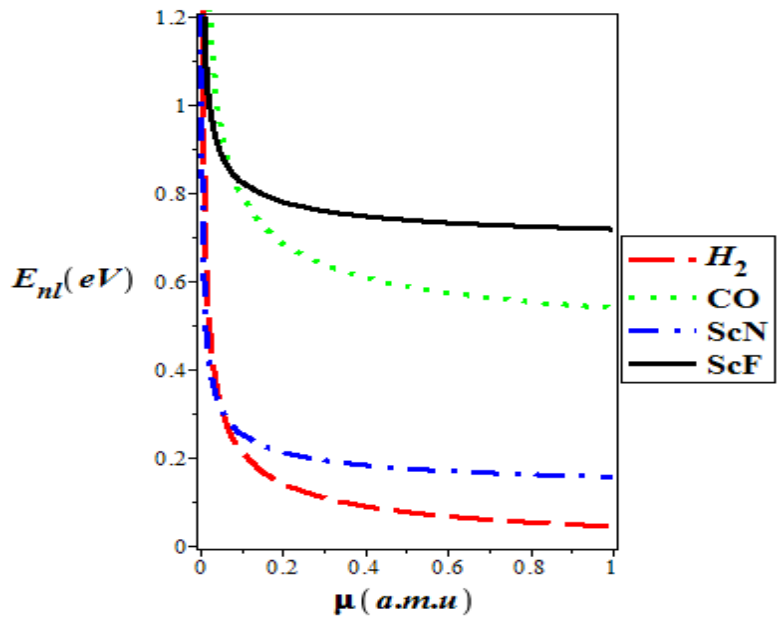

Figure 13: energy eigenvalues variation with the particle mass $\mu$ for various diatomic molecules

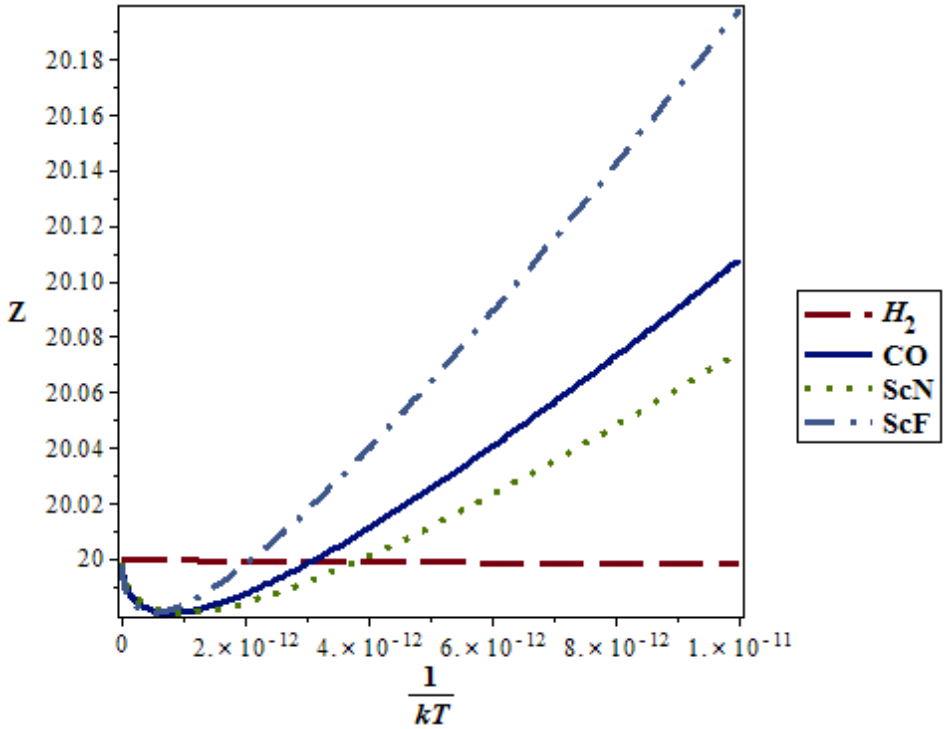

Figure 14: vibrational partition function variation with temperature for various diatomic molecules.

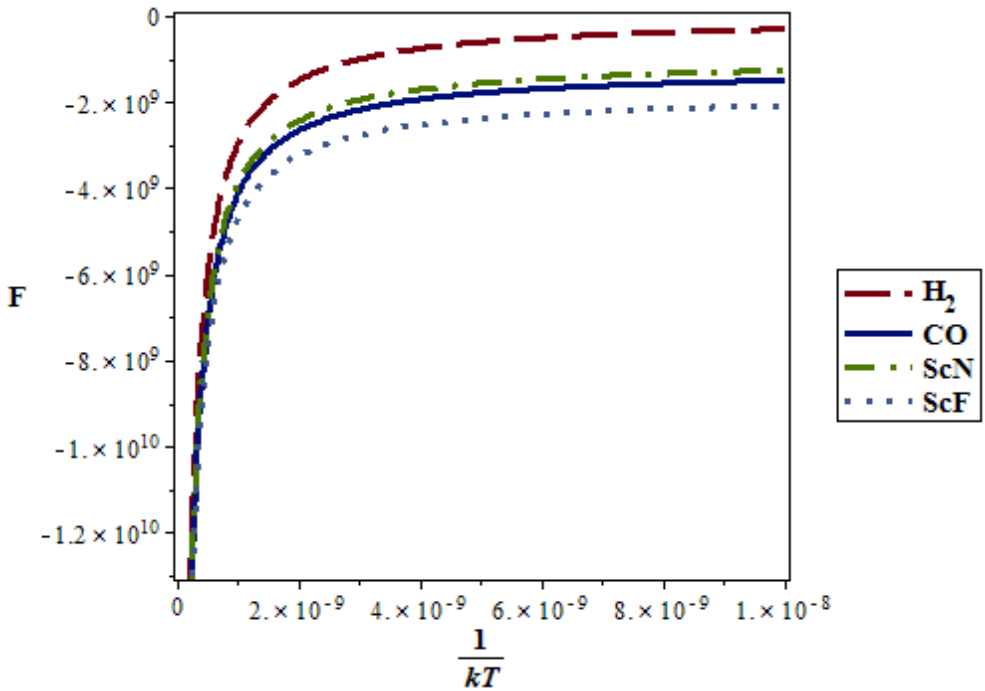

Figure 15: Vibrational free energy variation with temperature for various diatomic molecules.

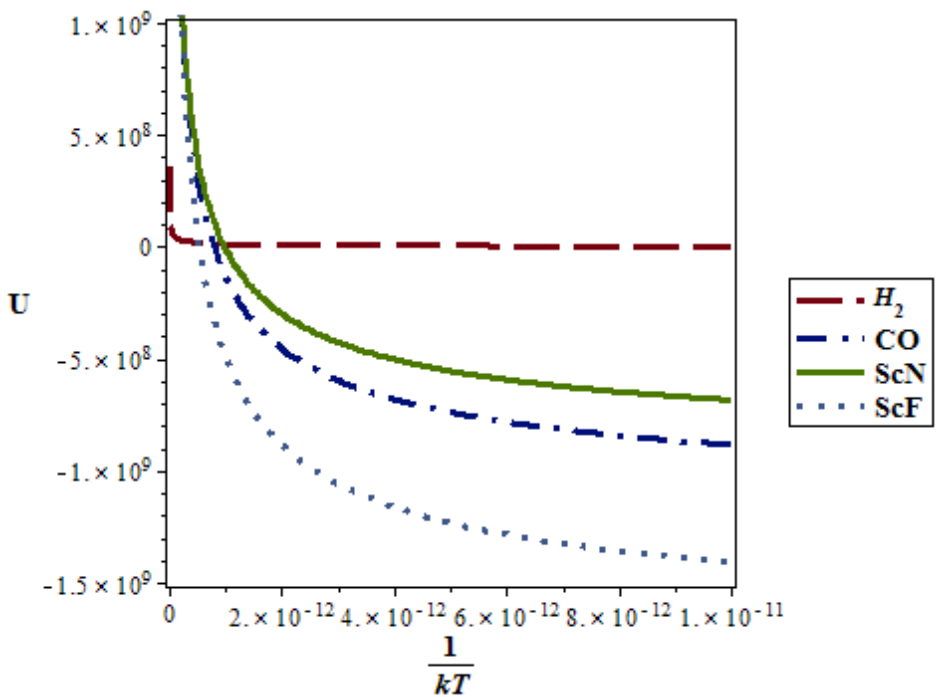

Figure 16: Vibrational mean energy variation with temperature for various diatomic molecules.

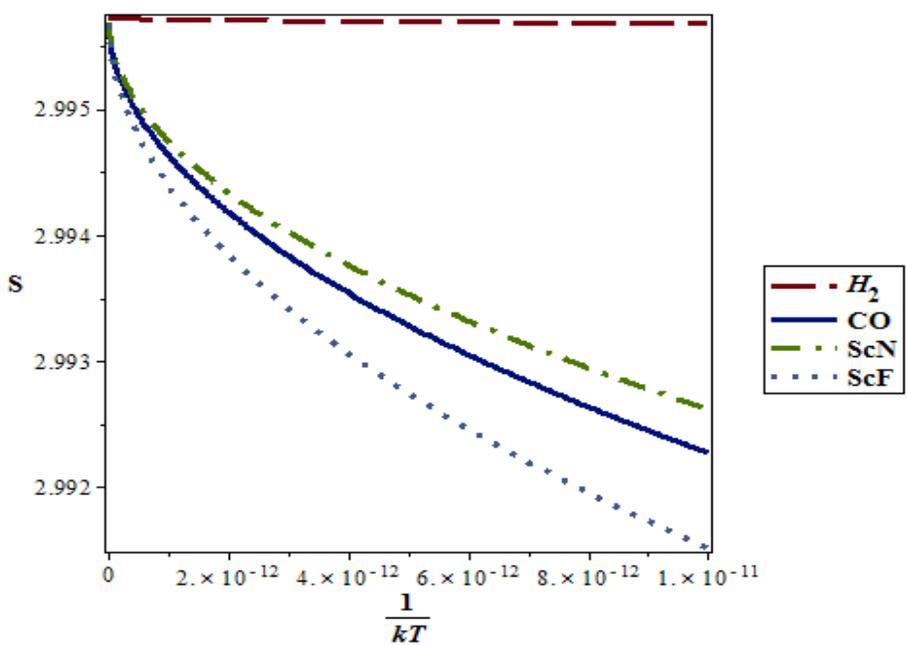

Figure 17: Vibrational entropy variation with temperature for various diatomic molecules.

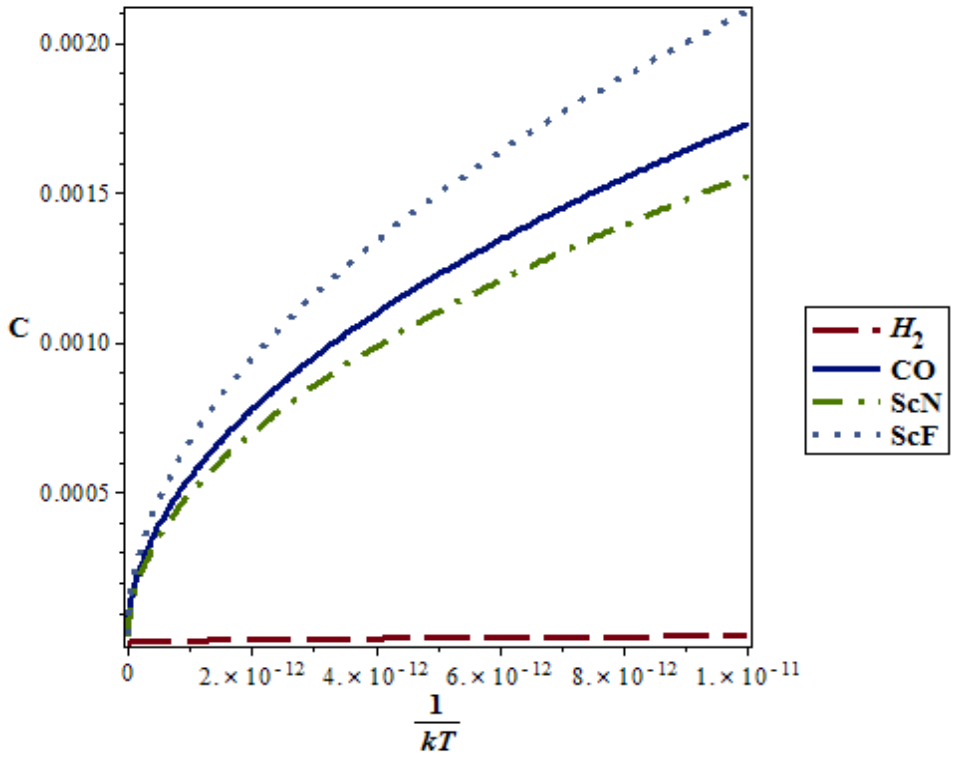

Figure 18: Vibrational specific heat capacity variation with temperature for various diatomic molecules.

Table 1; Spectroscopic parameters of the molecules used in this work

| Parameters | $H_2$ [31] | $CO$ [30] | $ScN$ [31] | $ScF$ [31] |
|---|---|---|---|---|
| $D_e \left(cm^{-1}\right)$ | 38266 | 90540 | 36778.88 | 47183.43 |
| $r_e$ (Å) | 0.7416 | 1.1283 | 1.768 | 1.794 |
| $\alpha$ (Å$^{-1}$) | 1.9426 | 2.294 | 1.5068 | 1.46102 |
| $\mu$ (a.m.u) | 0.50391 | 6.860672 | 10.68277 | 13.35894 |

Table 2. Comparison of energy eigenvalues (in *eV*) obtained by using Factorization Method with other methods for 2*p*, 3*p*, 3*d*, 4*d*, 4*f*, 5*p*, 5*d*, 5*f*, 5*g*, 6*p*,6*d*, 6*f* and 6*g* states in atomic units $\hbar = \mu = 1$ for $r_e = 0.4$ and $D_e = 15$.

| states | $\alpha$ | Present | NU[31] | AIM[66] | EXACT[60] | FAA[60] | LXC[67] |
|---|---|---|---|---|---|---|---|
| 2p | 0.05 | 7.860804501 | 7.860804493 | 7.860804467 | 7.862800 | 7.86060 | 7.86080 |
|  | 0.10 | 7.95330448 | 7.953304454 | 7.953304350 | 7.955370 | 7.95247 | 7.95330 |
|  | 0.15 | 8.045099826 | 8.04509987 | 8.045099635 | 8.047240 | 8.04322 | 8.04510 |
|  | 0.20 | 8.136203742 | 8.136203772 | 8.136203356 | 8.138420 | 8.13287 | 8.13620 |
|  | 0.25 | 8.226629166 | 8.226629167 | 8.226628516 | 8.228920 | 8.22142 | 8.22660 |
|  | 0.30 | 8.316389012 | 8.31638903 | 8.316388092 | 8.318740 | 8.30889 | 8.31639 |
| 3p | 0.05 | 10.99776305 | 10.99776305 | 10.99776302 | 10.999800 | 10.99760 | 10.99780 |
|  | 0.10 | 11.16256048 | 11.16256046 | 11.16256036 | 11.164700 | 11.16170 | 11.16260 |
|  | 0.15 | 11.32424869 | 11.32424872 | 11.32424848 | 11.326470 | 11.32240 | 11.32420 |
|  | 0.20 | 11.4828376 | 11.48283762 | 11.48283721 | 11.485130 | 11.47950 | 11.48280 |
|  | 0.25 | 11.63833667 | 11.63833667 | 11.63833602 | 11.640680 | 11.63310 | 11.63830 |
|  | 0.30 | 11.790755 | 11.79075502 | 11.79075408 | 11.675650 | 11.78330 | 11.79080 |
| 3d | 0.05 | 10.21598027 | 10.21598027 | 10.21598019 | 10.216510 | 10.21540 | 10.21598 |
|  | 0.10 | 10.3535395 | 10.35353947 | 10.35353916 | 10.354090 | 10.35100 | 10.35353 |
|  | 0.15 | 10.48935435 | 10.48935439 | 10.48935369 | 10.489920 | 10.48370 | 10.48935 |
|  | 0.20 | 10.62346372 | 10.62346374 | 10.62346249 | 10.624030 | 10.61350 | 10.62346 |
|  | 0.25 | 10.75590641 | 10.75590641 | 10.75590446 | 10.756450 | 10.74030 | 10.75591 |
|  | 0.30 | 10.88672148 | 10.88672151 | 10.88671869 | 10.887190 | 10.86420 | 10.88672 |
| 4p | 0.05 | 12.49760242 | 12.49760242 | 12.4976024 | 12.499200 | 12.49740 | 12.49760 |
|  | 0.10 | 12.69679606 | 12.69679604 | 12.69679594 | 12.698510 | 12.69600 | 12.69680 |
|  | 0.15 | 12.88834811 | 12.88834813 | 12.8883479 | 12.890100 | 12.88650 | 12.88835 |
|  | 0.20 | 13.07224461 | 13.07224462 | 13.0722442 | 13.074000 | 13.06890 | 13.07224 |
|  | 0.25 | 13.24847044 | 13.24847044 | 13.24846979 | 13.250100 | 13.24330 | 13.24847 |
| 4d | 0.05 | 12.09829027 | 12.09829027 | 12.09829019 | 12.098900 | 12.09770 | 12.09829 |
|  | 0.10 | 12.28500944 | 12.28500942 | 12.2850091 | 12.285700 | 12.28250 | 12.28501 |
|  | 0.15 | 12.46641934 | 12.46641937 | 12.46641867 | 12.467150 | 12.46080 | 12.46642 |
|  | 0.20 | 12.64256754 | 12.64256756 | 12.64256631 | 12.643240 | 12.63260 | 12.64257 |
| 4f | 0.05 | 11.82078624 | 11.82078623 | 11.82078608 | 11.820900 | 11.81950 | 11.82079 |
|  | 0.10 | 11.99796123 | 11.99796121 | 11.99796058 | 11.998100 | 11.99300 | 11.99796 |
|  | 0.15 | 12.17169657 | 12.17169661 | 12.1716952 | 12.171800 | 12.16040 | 12.17170 |
|  | 0.20 | 12.34207215 | 12.34207217 | 12.34206967 | 12.342100 | 12.32210 | 12.34207 |
| 5p | 0.10 | 13.54214251 | 13.5421425 | 13.5421424 | 13.543400 | 13.54130 | 13.54214 |
|  | 0.20 | 13.92898632 | 13.92898633 | 13.92898591 | 13.930100 | 13.92570 | 13.92999 |
| 5d | 0.10 | 13.30679692 | 13.3067969 | 13.30679659 | 13.307500 | 13.30430 | 13.30680 |
|  | 0.20 | 13.69266394 | 13.69266395 | 13.6926627 | 13.693100 | 13.68270 | 13.69266 |
| 5f | 0.10 | 13.14759773 | 13.14759771 | 13.14759709 | 13.147800 | 13.14260 | 13.14760 |
|  | 0.20 | 13.53344224 | 13.53344225 | 13.53343975 | 13.533300 | 13.51340 | 13.53344 |
| 5g | 0.10 | 13.03797622 | 13.03797622 | 13.03797516 | 13.037900 | 13.02960 | 13.03798 |
| | 0.20 | 13.42711264 | 13.42711266 | 13.4271085 | 13.426670 | 13.39380 | 13.42711 |
| 6p | 0.10 | 14.05208861 | 14.05208861 | 14.0520885 | 14.053000 | 14.05130 | 14.05209 |
| 6d | 0.10 | 13.90704847 | 13.90704846 | 13.90704815 | 13.907500 | 13.90450 | 13.90705 |
| 6f | 0.10 | 13.81118996 | 13.81118995 | 13.81118932 | 13.811300 | 13.80620 | 13.81119 |
| 6g | 0.10 | 13.74661284 | 13.74661283 | 13.74661179 | 13.746600 | 13.73830 | 13.74361 |

Table 3. Comparison of energy eigenvalues (in $eV$) obtained by using Factorization Method with other methods for 2p, 3p, 3d, 4d, 4f, 5p, 5d, 5f, 5g, 6p, 6d, 6f and 6g states in atomic units $\hbar = \mu = 1$ for $r_e = 0.8$ and $D_e = 15$.

| States | $\alpha$ | Present | NU[31] | AIM[66] | EXACT[60] | APPROXIMATE[60] | LXC[67] |
|---|---|---|---|---|---|---|---|
| 2p | 0.05 | 4.140887280 | 4.140887263 | 4.140887237 | 4.14208 | 4.14068 | 4.140887 |
|  | 0.10 | 4.219180100 | 4.219180128 | 4.219180023 | 4.22040 | 4.21835 | 4.219180 |
|  | 0.15 | 4.297393180 | 4.297393199 | 4.297392964 | 4.29870 | 4.29552 | 4.297393 |
|  | 0.20 | 4.375546500 | 4.375546508 | 4.375546092 | 4.37690 | 4.37221 | 4.375547 |
|  | 0.25 | 4.453659650 | 4.453659654 | 4.453659003 | 4.45510 | 4.44845 | 4.653660 |
|  | 0.30 | 4.531751800 | 4.531751791 | 4.531750853 | 4.53320 | 4.52425 | 4.531752 |
| 3p | 0.05 | 7.532791581 | 7.532791561 | 7.532791535 | 7.53500 | 7.53258 | 7.532792 |
|  | 0.10 | 7.724764250 | 7.724764274 | 7.724764169 | 7.72710 | 7.72393 | 7.724764 |
|  | 0.15 | 7.915178638 | 7.915178655 | 7.915178421 | 7.91770 | 7.9133 | 7.915179 |
|  | 0.20 | 8.104040620 | 8.104040627 | 8.104040211 | 8.10660 | 8.10071 | 8.104041 |
|  | 0.25 | 8.291354170 | 8.291354169 | 8.291353518 | 8.29410 | 8.28615 | 8.291354 |
|  | 0.30 | 8.477121312 | 8.477121312 | 8.477120373 | 8.47990 | 8.46962 | 8.477121 |
| 3d | 0.05 | 5.739751255 | 5.739751228 | 5.73975115 | 5.74040 | 5.73913 | 5.739751 |
|  | 0.10 | 5.845770258 | 5.845770281 | 5.845769968 | 5.84650 | 5.84327 | 5.845770 |
|  | 0.15 | 5.950678118 | 5.950678133 | 5.95067743 | 5.95150 | 5.94505 | 5.950678 |
|  | 0.20 | 6.054533595 | 6.054533598 | 6.054532348 | 6.05530 | 6.04453 | 6.054534 |
|  | 0.25 | 6.157395320 | 6.157395321 | 6.157393368 | 6.15820 | 6.14177 | 6.157395 |
|  | 0.30 | 6.259321760 | 6.259321745 | 6.259318933 | 6.26010 | 6.23682 | 6.259322 |
| 4p | 0.05 | 9.613013105 | 9.613013087 | 9.613013061 | 9.61560 | 9.6128 | 9.613013 |
|  | 0.10 | 9.883523678 | 9.883523698 | 9.883523594 | 9.88620 | 9.88269 | 9.983524 |
|  | 0.15 | 10.148555700 | 10.14855572 | 10.14855549 | 10.15140 | 10.1467 | 10.148560 |
|  | 0.20 | 10.408057740 | 10.40805775 | 10.40805734 | 10.41110 | 10.4047 | 10.408060 |
|  | 0.25 | 10.661973880 | 10.66197388 | 10.66197323 | 10.66500 | 10.6568 | 10.661970 |
| 4d | 0.05 | 8.493343511 | 8.493343486 | 8.493343408 | 8.49480 | 8.49272 | 8.493344 |
|  | 0.10 | 8.707110961 | 8.707110984 | 8.707110672 | 8.70870 | 8.70461 | 8.707111 |
|  | 0.15 | 8.917807581 | 8.917807599 | 8.917806896 | 8.91940 | 8.91218 | 8.917808 |
|  | 0.20 | 9.125505091 | 9.125505093 | 9.125503844 | 9.12720 | 9.11551 | 9.125505 |
| 4f | 0.05 | 7.434705841 | 7.434705812 | 7.434705654 | 7.43510 | 7.43346 | 7.434706 |
|  | 0.10 | 7.586418776 | 7.586418806 | 7.586418181 | 7.58680 | 7.58142 | 7.586419 |
|  | 0.15 | 7.735732245 | 7.735732273 | 7.735730867 | 7.73610 | 7.72448 | 7.735732 |
|  | 0.20 | 7.882757509 | 7.882757512 | 7.882755012 | 7.88310 | 7.86276 | 7.882758 |
| 5p | 0.10 | 11.302072420 | 11.30207244 | 11.30207233 | 11.30470 | 11.3012 | 11.302070 |
|  | 0.20 | 11.913223740 | 11.91322375 | 11.91322333 | 11.91610 | 11.9099 | 11.913220 |
| 5d | 0.10 | 10.520086060 | 10.52008608 | 10.52008576 | 10.52190 | 10.5176 | 10.520090 |
|  | 0.20 | 11.069371600 | 11.06937161 | 11.06937036 | 11.07130 | 11.0594 | 11.069370 |
| 5f | 0.10 | 9.796658009 | 9.796658033 | 9.796657408 | 9.79750 | 9.79166 | 9.796658 |
|  | 0.20 | 10.273037060 | 10.27303707 | 10.27303457 | 10.27380 | 10.253 | 10.273040 |
| 5g | 0.10 | 9.152223325 | 9.152223355 | 9.152222313 | 9.15240 | 9.14389 | 9.522230 |
| | 0.20 | 9.552869470 | 9.552869479 | 9.552865312 | 9.55280 | 9.51954 | 9.552869 |
| 6p | 0.10 | 12.279799200 | 12.27979921 | 12.27979911 | 12.28220 | 12.279 | 12.279800 |
| 6d | 0.10 | 11.736438620 | 11.73643864 | 11.73643833 | 11.73830 | 11.7339 | 11.736440 |
| 6f | 0.10 | 11.244814900 | 11.24481492 | 11.2448143 | 11.24590 | 11.2398 | 11.244810 |
| 6g | 0.10 | 10.815332260 | 10.81533228 | 10.81533124 | 10.81580 | 10.807 | 10.815330 |

**Table 4:** The bound state energies $E_{n\ell}(eV)$ for $H_2$, $CO$, $ScN$ and $ScF$ molecules for different values of the vibrational $n$ and rotational $\ell$ quantum numbers of the Deng–Fan-Eckart potential

| $n$ | $\ell$ | $H_2$ | $CO$ | $ScN$ | $ScF$ |
|---|---|---|---|---|---|
| 0 | 0 | 0.3161089 | 0.3653484 | 0.5678234 | 0.7807499 |
| 0 | 1 | 0.3304972 | 0.3658917 | 0.5679634 | 0.7808573 |
| 1 | 0 | 0.9626475 | 0.6487759 | 0.6573524 | 0.8685782 |
| 1 | 1 | 0.9759181 | 0.6493153 | 0.6574917 | 0.8686851 |
| 2 | 0 | 1.5466492 | 0.9281700 | 0.7457977 | 0.9555848 |
| 2 | 1 | 1.5588663 | 0.9287056 | 0.7459361 | 0.9556913 |
| 2 | 2 | 1.5832386 | 0.9297766 | 0.7462130 | 0.9559042 |
| 3 | 0 | 2.0716668 | 1.2035434 | 0.8331616 | 1.0417712 |
| 3 | 1 | 2.0828895 | 1.2040751 | 0.8332993 | 1.0418772 |
| 3 | 2 | 2.1052768 | 1.2051383 | 0.8335745 | 1.0420891 |
| 3 | 3 | 2.1387131 | 1.2067333 | 0.8339876 | 1.0424072 |
| 4 | 0 | 2.5409692 | 1.4749081 | 0.9194467 | 1.1271390 |
| 4 | 1 | 2.5512519 | 1.4754360 | 0.9195836 | 1.1272445 |
| 4 | 2 | 2.5717628 | 1.4764914 | 0.9198573 | 1.1274555 |
| 4 | 3 | 2.6023934 | 1.4780750 | 0.9202679 | 1.1277721 |
| 4 | 4 | 2.6429822 | 1.4801861 | 0.9208153 | 1.1281941 |
| 5 | 0 | 2.9575685 | 1.7422767 | 1.0046555 | 1.2116896 |
| 5 | 1 | 2.9669614 | 1.7428007 | 1.0047916 | 1.2117947 |
| 5 | 2 | 2.9856959 | 1.7438485 | 1.0050637 | 1.2120047 |
| 5 | 3 | 3.0136703 | 1.7454204 | 1.0054720 | 1.2123199 |
| 5 | 4 | 3.0507328 | 1.7475162 | 1.0060162 | 1.2127400 |
| 5 | 5 | 3.0966829 | 1.7501360 | 1.0066966 | 1.2132652 |